\documentclass[prl,showpacs,superscriptaddress,twocolumn]{revtex4}

\usepackage{graphicx}
\usepackage{amsmath}
\usepackage{amssymb}
\usepackage{mathrsfs}

\usepackage[british]{babel}

\begin{document}

\title{In vivo anomalous diffusion and weak ergodicity breaking of lipid
granules}

\author{Jae-Hyung Jeon}
\affiliation{Physics Department T30g, Technical University of Munich,
85747 Garching, Germany}
\author{Vincent Tejedor}
\affiliation{Physique Th{\'e}orique de la mati{\'e}re condens{\'e}e,
Universit{\'e} Pierre et Marie Curie, 4 place Jussieu, 75252 Paris, France}
\author{Stas Burov}
\affiliation{Physics Department, Bar-Ilan University, Ramat Gan 52900, Israel}
\author{Eli Barkai}
\affiliation{Physics Department, Bar-Ilan University, Ramat Gan 52900, Israel}
\author{Christine Selhuber-Unkel}
\affiliation{Niels Bohr Institute, Blegdamsvej 17, 2100 K{\o}benhavn {\O},
Denmark}
\affiliation{Institute for Materials Science, University of Kiel,
Kaiserstra{\ss}e 2, 24143 Kiel, Germany}
\author{Kirstine Berg-S{\o}rensen}
\affiliation{Physics Department, Technical University of Denmark,
2800 Kongens Lyngby, Denmark}
\author{Lene Oddershede}
\affiliation{Niels Bohr Institute, Blegdamsvej 17, 2100 K{\o}benhavn {\O},
Denmark}
\author{Ralf Metzler}
\affiliation{Physics Department T30g, Technical University of Munich,
85747 Garching, Germany}

\date{\today}

\begin{abstract}
Combining extensive single particle tracking microscopy data of endogenous
lipid granules in living fission yeast cells with analytical
results we show evidence for anomalous diffusion and weak ergodicity
breaking. Namely we demonstrate that at short times the granules perform
subdiffusion according to the laws of continuous time random walk theory.
The associated violation of ergodicity leads to a characteristic turnover
between two scaling regimes of the time averaged mean squared displacement.
At longer times the granule motion is consistent with fractional Brownian
motion.
\end{abstract}


\maketitle

Crowded colloidal systems of hard and soft core particles
display a rich physical behavior including dynamic
arrest, non-exponential relaxation, and anomalous diffusion
\cite{mattsson}. In biological cells the cytoplasm is a crowded soft core
colloid made up of large biopolymers such as ribosomes, proteins, or RNA,
occupying volume fractions of 34\% or above \cite{minton,mcguffee}, amidst
small particles such as water, ions, and lower mass fraction biopolymers.
Such crowding significantly impacts cellular biochemical reactions
\cite{minton} and effects subdiffusion of the form
\cite{report,REM1}
\begin{equation}
\label{msd}
\langle\mathbf{r}^2(t)\rangle\propto K_{\alpha}t^{\alpha},\quad0<\alpha<1,
\end{equation}
of larger molecules or tracers in living cells \cite{saxton,golding,guigas,%
garini,seisenhuber,vercammen,elbaum,lene,weiss1,weber,mcguffee,christine}.
While normal diffusion, by
virtue of the central limit theorem, is characterized by the universal Gaussian
probability density function and therefore uniquely determined by the first
and second moments \cite{hughes}, anomalous diffusion of the form (\ref{msd})
is non-universal and may be caused by different stochastic mechanisms. These
would give rise to vastly different behavior for
diffusional mixing, diffusion-limited reactions, signaling, or regulatory
processes. To better understand cellular dynamics, knowledge of
the underlying stochastic mechanism is thus imperative.

Here we report experimental evidence from extensive single trajectory time
series of lipid granule motion in \emph{Schizosaccharomyces pombe\/}
(\emph{S.~pombe}) fission yeast cells obtained from tracking with
optical tweezers (resolving $10^{-6}$ sec to $1$ sec) and video microscopy
($10^{-2}$ sec to 100 sec). Using complementary analysis tools we demonstrate
that at short times the data are described best by continuous time random
walk (CTRW) subdiffusion, revealing pronounced features of weak ergodicity
breaking in the time averaged mean squared displacement. At longer
times the stochastic mechanism is closest to subdiffusive fractional Brownian
motion (FBM). The time scales over which this anomalous behavior persists is
relevant for biological processes occurring in the cell.
Anomalous diffusion may indeed be a good strategy for cellular signaling
and reactions \cite{golding,guigas}.

CTRW and FBM both effect anomalous diffusion of the type (\ref{msd})
\cite{REM}. Subdiffusive CTRWs are random walks with finite
variance $\langle\delta x^2\rangle$ of jump lengths, while the waiting
times between successive jumps are drawn from a density $\psi(t)\simeq\tau^{
\alpha}/t^{1+\alpha}$ with diverging characteristic time \cite{report,hughes}.
Such scale-free behavior results from multiple trapping events in, e.g.,
comb-like structures \cite{havlin} or random energy landscapes \cite{rel}.
Power-law waiting time distributions
were also identified for tracer motion in reconstituted
actin networks \cite{weitz}. Subdiffusive FBM is a random process driven by
Gaussian noise $\xi$ with long-range correlations, $\langle\xi(0)\xi(t)\rangle
\simeq\alpha(\alpha-1)t^{\alpha-2}$ \cite{mandelbrot}, and it is related to
fractional Langevin equations \cite{fle}. The finite
characteristic time scales associated with FBM contrast the ageing property
of the subdiffusive CTRW processes [Supplementary Material (SM)].

Single particle tracking microscopy has become a standard tool to probe the
motion of individual tracers, especially inside living cells, and provides
valuable insights into cellular dynamics \cite{saxton}. The recorded time
series $\mathbf{r}(t)$ are analyzed in terms of the time
averaged mean squared displacement as function of the lag time $\Delta$,
\begin{equation}
\label{tamsd}
\overline{\delta^2(\Delta,T)}=\frac{1}{T-\Delta}\int_0^{T-\Delta}\Big[
\mathbf{r}(t+\Delta)-\mathbf{r}(t)\Big]^2dt,
\end{equation}
where $T$ is the total measurement time. CTRW and FBM subdiffusion show
markedly different behavior of $\overline{\delta^2(\Delta,T)}$
\cite{he,lub,stas,igor1}. Subdiffusion of the form $\overline{
\delta^2}\simeq\Delta^{\alpha}$ was found in several tracking experiments
\emph{in vivo\/} \cite{golding,garini,seisenhuber,elbaum,lene,christine,weber}.
Crowding-induced subdiffusion was also observed in controlled environments,
e.g., in concentrated protein and dextran solutions
\cite{weiss,weiss1,banks,pan}. While FBM was proposed as stochastic mechanism
in some of these systems \cite{weber,weiss1,marcin} CTRW subdiffusion
has not yet been identified. Here we demonstrate that the short time motion of
lipid granules in \emph{S.pombe\/} cells follows the laws of CTRW
subdiffusion and features weak ergodicity breaking.

\begin{figure}
\includegraphics[width=8cm]{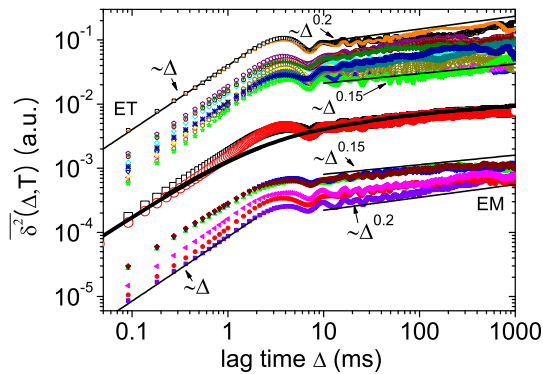}
\caption{Time averaged mean squared displacement $\overline{\delta^2}$ from
individual trajectories of lipid granules in \emph{S.pombe} in Early Mitotic
cells (lower curves) and in Early Telophase (upper curves), measured by optical
tweezers \cite{REM2}. A distinct turnover from $\overline{\delta^2}\simeq\Delta$
to $\simeq\Delta^{\beta}$ ($\beta\approx0.10\ldots0.20$) occurs. The two thick
lines in the middle show the averages of the ET (red $\bigcirc$) and EM
(black $\square$) data sets. The overlaid thick black line is the result of
CTRW simulations in an harmonic potential. ET versus EM curves are shifted
vertically.}
\label{shorttime}
\end{figure}

\emph{Short time behavior.} Granule motion was recorded in
an optical tweezers setup (SM). In the experiment the
trap is initially centered onto the granule such that no force is exerted. When
the granule starts to move away from the trap center it experiences a restoring
Hookean force \cite{trap}. The measured $\overline{\delta^2(\Delta)}$ curves
from two different cell stages are shown in Fig.~\ref{shorttime} \cite{REM2},
along with a typical sample trajectory in Fig.~5a (SM). In Fig.~\ref{shorttime}
a distinct turnover is observed from an initial linear growth $\overline{\delta
^2}\simeq\Delta$ to a power-law behavior $\overline{\delta^2}\simeq\Delta^{
\beta}$ with $\beta\approx0.15\ldots0.20$.

The size of the lipid granules is about 300 nm. In addition to the amplitude
scatter between different trajectories expected from CTRW theory \cite{he,stas},
the fluctuations of the data in Fig.~\ref{shorttime} are due to natural granule
size variations and different optical conditions for each trajectory. We note
that oscillations around the turnover may arise for subdiffusion in an
underdamped medium \cite{stas_osc}.

Evaluating extensive data for the granule motion analogous to the data presented
here by \emph{ensemble\/} averages, anomalous diffusion with
$\alpha\approx0.80\ldots0.85$ yields in a range of 0.1 to 3 msec
\cite{christine}.
In contrast, in Fig.~\ref{shorttime} the initial behavior of $\overline{
\delta^2}$ corresponding to this time range does not exhibit any apparent
anomaly but scales like $\overline{\delta^2}\simeq\Delta^1$. At longer lag
time $\Delta$ due to the
trap force one would expect $\overline{\delta^2}$ to saturate to a
stationary thermal value; instead,
the regime $\overline{\delta^2}\simeq\Delta^{\beta}$ appears. This behavior
is consistently
observed in different cell stages (Fig.~\ref{shorttime}).

Such a peculiar behavior is fully consistent with CTRW subdiffusion: For free
motion
one finds $\overline{\delta^2}\simeq\Delta/T^{1-\alpha}$, whose $\Delta$
scaling is independent of $\alpha$, while
the corresponding ensemble average follows Eq.~(\ref{msd}) \cite{he,lub}. Under
confinement a turnover to the power-law $\overline{\delta^2}\simeq(\Delta/
T)^{1-\alpha}$ occurs \cite{stas,igor1}.
This second power-law regime is terminated when
$\Delta$ approaches the total measurement time $T$, causing a dip in $\overline{
\delta^2}$ back to the plateau of the ensemble average. The observed
characteristic
turnover behavior is intimately connected to CTRW ageing and ergodicity breaking
\cite{ageing,web}. Corralled motion \cite{corral} could not explain the
observed behahior $\overline{\delta^2}\simeq\Delta$ turning over to
$\overline{\delta^2}\simeq\Delta^{\beta}$.

The following features unanimously point toward CTRW subdiffusion as the
stochastic mechanism for the granule motion at short times: (i) The time
average $\overline{\delta^2}$ initially scales linearly with $\Delta$, albeit
the ensemble average shows subdiffusion of the type (\ref{msd}) in comparable
time ranges. (ii) At longer times a turnover to the power-law $\overline{\delta
^2}\simeq\Delta^{\beta}$ occurs instead of the convergence to a plateau, which
would necessarily occur for an ergodic process; the anomalous diffusion exponent
$\alpha\approx0.80\ldots0.85$ observed in the ensemble average is consistent
with the exponents $\beta\approx0.15\ldots0.20$ observed in Fig.~\ref{shorttime}
based on the relation $\beta=1-\alpha$ \cite{stas,igor1}, as well as the slopes
of the long time data (see
below). (iii) The specific form of $\overline{\delta^2}$ of the data nicely
coincides with simulations results for a subdiffusive CTRW in an harmonic
potential (Fig.~\ref{shorttime}). These observations prove that the granule
motion in fact exhibits weak ergodicity breaking associated with subdiffusive
CTRW processes. This is the central finding of this work.

We rule out the possibility that the initial CTRW simply turns
over to much slower diffusion with $\alpha\approx0.15$ $\ldots0.20$, as the
overlapping video tracking data shows subdiffusion with $\alpha\approx0.8$ or
slightly below. Also, CTRW or FBM trajectories with $\alpha=0.20$ would cause
a quite extreme stalling or antipersistence, respectively,
which is inconsistent with the recorded time series [Fig.~5 (SM)].

\begin{figure}
\includegraphics[width=8cm]{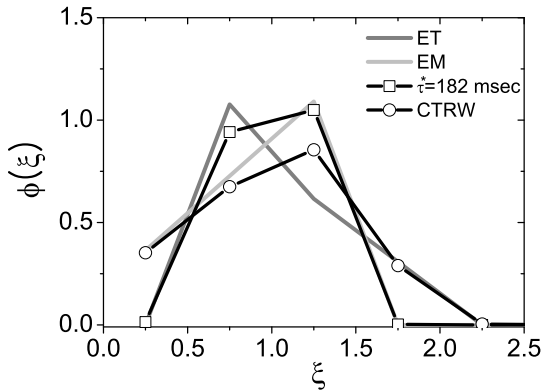}
\caption{Distribution $\phi(\xi)$ of the relative deviations $\xi=\overline{
\delta^2}/\langle\overline{\delta^2}\rangle$ of $\overline{\delta^2}$
obtained from averaging the data of Fig.~\ref{shorttime} (Early Telophase,
Early Mitosis). Simulations results are shown based on Eq.~(\ref{wtd}) with
$\tau^*=182$ msec ($\square$), and for a CTRW with $\langle t\rangle=\infty$
($\bigcirc$), with parameters $\alpha=0.85$, $T=3.0$ sec, $\tau=0.04545$
msec, and $\Delta=0.04545$ msec. See Fig.~7 (SM).}
\label{phi_stage3}
\end{figure}

What results yield from complimentary criteria? (i) Evaluating $\overline{
\delta^2(\Delta,T)}$ as function of the total measurement time $T$, no ageing
is observed [Fig.~6 (SM)], contrasting the scaling $\overline{\delta^2(T)}
\simeq T^{\alpha-1}$ predicted for CTRW subdiffusion with diverging mean
waiting time $\langle t\rangle$ \cite{he,lub,stas}. (ii) Determining the
distribution $\phi(\xi)$ of the deviations of the relative time averaged mean
squared displacement $\xi=\overline{\delta^2}/\langle\overline{\delta^2}
\rangle$ around the ensemble mean ($\xi=1$) we find the bell-shaped form of
Fig.~\ref{phi_stage3} with $\phi(0)\approx0$. This form deviates from the
predicted shape for CTRW subdiffusion with $\langle t\rangle=\infty$
where $\phi(0)\neq0$ \cite{he}, see Fig.~\ref{phi_stage3}.

\begin{figure}
\includegraphics[width=8cm]{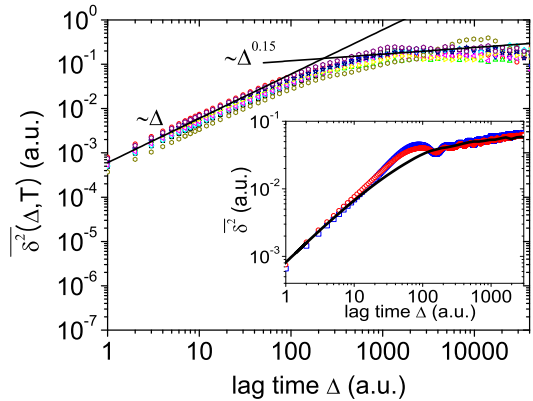}
\caption{Time averaged mean squared displacement for a CTRW with waiting time
cutoff (\ref{wtd}) in an harmonic potential.
Inset: Comparison of the average behavior of the simulated
trajectories (---) and the average of the experimental data from
Fig.~\ref{shorttime} ($\bigcirc,\square$). Parameters as in
Fig.~\ref{phi_stage3}.}
\label{cutoff_tamsd}
\end{figure}

These seemingly conflicting observations can in fact be reconciled when we
consider a power-law waiting time distribution with a cutoff effective at
times $t\gg\tau^*$,
\begin{equation}
\label{wtd}
\psi(t)=\frac{d}{dt}\left[1-\frac{\tau^{\alpha}}{(\tau+t)^{\alpha}}\exp
\left(-\frac{t}{\tau^*}\right)\right].
\end{equation}
Physically such cutoffs occur naturally in finite systems, corresponding to,
e.g., a maximal well depth in a random energy landscape. In free space
Eq.~(\ref{wtd}) produces initial subdiffusion $\langle\mathbf{r}^2(t)\rangle
\simeq t^{\alpha}$, turning over to $\langle\mathbf{r}^2(t)\rangle\simeq t$ at
$t\gg\tau^*$. As shown in Fig.~\ref{cutoff_tamsd} the cutoff-waiting time
distribution (\ref{wtd}) reproduces well the experimental behavior from
Fig.~\ref{shorttime}. The scatter distribution $\phi(\xi)$
[Figs.~\ref{phi_stage3}, 7 (SM)] shows excellent agreement, and the ageing
plot [Fig.~6 (SM)] is consistent with the data. Here we chose the cutoff
time $\tau^*=182$ msec, and $\tau=0.04545$ msec.

Collecting all results we conclude that at short times the granule motion is
described by CTRW subdiffusion with truncated power-law waiting time
distribution (\ref{wtd}). While the cutoff ensures that the typical scaling of
$\overline{\delta^2(\Delta,T)}$ still exhibits the weak ergodicity
breaking features, for appropriate choice of the cutoff time it is consistent
with the absence of ageing effects, i.e., $\overline{\delta^2(T)}\simeq T^0$,
and the observed form for $\phi(\xi)$.

\begin{figure}
\includegraphics[width=7.2cm]{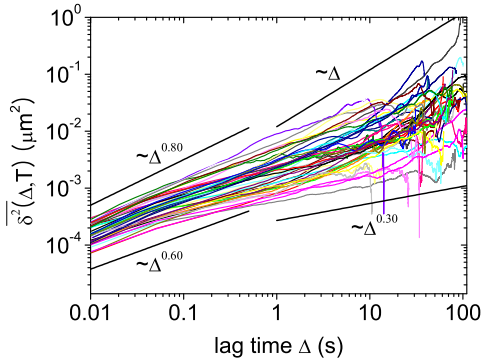}
\caption{Time averaged mean squared displacement of granules in \emph{S.pombe}
(cells in interphase) from video tracking data (SM). An initial slope around
$\alpha\approx0.8$ is found,
turning over to a gentler slope. Several trajectories later exhibit $\alpha
\approx1.0$.}
\label{longtime_tamsd}
\end{figure}

\emph{At longer times\/} the motion of the lipid granules was recorded by
video particle tracking. Fig.~\ref{longtime_tamsd}
shows the time averaged mean squared displacement. Initially the slope is
around $\alpha\approx0.8$ or slightly below, consistent with the short time
data. Several of the curves turn to a gentler slope at around 100 msec, some
curves eventually switch
to normal diffusion ($\alpha=1$) at 1 sec. Significant deviations
are observed, which, for a living systems, is not surprising. Both the granule
size and the materials properties of the cellular environment may change on
these time scales
(depolymerization/repolymerization of the cytoskeleton etc). As shown in
Fig.~8 (SM) the distribution of $\phi(\xi)$ is bell-shaped; no ageing is
observed.
Analysis of the moment ratios for both normal moments and the mean maximal
excursion statistics (see Ref.~\cite{vincent} for details) are consistent with
FBM for the range of $\alpha\approx0.80\ldots0.85$ [Fig.~9 (SM)], as are
typical antipersistent trajectories [Fig.~10 (SM)].
Moreover, the scaling exponents of the mean maximal excursion second moment
are consistently above those of the corresponding regular second moments,
as predicted for FBM \cite{vincent}. The velocity
autocorrelation of the data is consistent with both FBM and CTRW [Fig.~11 (SM)],
and therefore is not conclusive. The p-variation method \cite{marcin} was
not conclusive.

Our analysis of extensive single particle tracking data of lipid granules in
\emph{S.pombe\/} cells demonstrates that at shorter times in the msec range
the motion displays significant effects of weak ergodicity breaking both in
the limit of free motion and in the presence of the restoring trap force.
Concurrently no ageing occurs and the distribution of
time averages $\phi(\xi)$ is bell-shaped around $\xi=1$. We showed that these
features are consistent with a CTRW process with truncated power-law
waiting time distribution. At longer times the motion is best described by
subdiffusive FBM, although a conclusive statement in this time range is more
difficult due to the fact that cellular processes appear to be superimposed on
the motion. Physically, the CTRW-like motion may be associated with the
interaction between granules and the semiflexible filaments of the
cytoskeleton similar to the observations in Ref.~\cite{weitz}.
While the behavior becomes more erratic in the long time
data a turnover to a gentler slope is observed before a final increase to
normal diffusion. This behavior may be connected to the viscoelastic properties
of the complex cellular environment. The identification of FBM as stochastic
mechanism is consistent with conclusions in Refs.~\cite{weber,weiss1}.

Lipid granules provide a natural, inert tracer to explore the diffusion
properties inside living cells.
Subdiffusion of large biopolymers in the cell supports the emerging, more
local picture of cellular transport and regulation \cite{mirny}. CTRW gives
rise to a dynamic spatial localization \cite{ageing,garini}, while FBM
provides a more compact spatial exploration and thus increases the local
encounter probability \cite{guigas}. Identifying these different subdiffusion
mechanisms in cells can give new insights into \emph{in vivo\/} molecular
processes at different timescales.

\acknowledgments

This project was supported by the Deutsche For\-schungs\-ge\-meinschaft (DFG),
the CompInt Graduate School, and the Israeli Science Foundation.


\begin{thebibliography}{99}
\bibitem{REM2} The voltage signal from the photodiode is directly proportional
to the particle position relative to the trap center. Imprecise knowledge of
the optical properties in the cell hampers a precise conversion to absolute
length.
\end{thebibliography}
\end{document}